\documentclass[fleqn,usenatbib]{mnras}

\usepackage{newtxtext,newtxmath}

\usepackage[T1]{fontenc}

\DeclareRobustCommand{\VAN}[3]{#2}
\let\VANthebibliography\thebibliography
\def\thebibliography{\DeclareRobustCommand{\VAN}[3]{##3}\VANthebibliography}

\usepackage{graphicx}	
\usepackage{amsmath}	
\usepackage[normalem]{ulem} 
\usepackage{soul}
\usepackage{CJKutf8}

\newcommand{\tktext}[1]{\begin{CJK}{UTF8}{mj}#1\end{CJK}}

\title[pressure bump]{Revisiting planetesimal formation from mm-sized grains in pressure bumps}

\author[T.S Tanvir et al.]{
Tabassum S Tanvir,$^{1}$\thanks{E-mail: ttanvir@iastate.edu (TST)}
Jacob B Simon,$^{1}$
Daniel Carrera$^{2}$
and Jeonghoon Lim (\tktext{임정훈})$^{1,3,4}$
\\
$^{1}$Department of Physics and Astronomy, Iowa State University, Ames, IA 50010, \\
$^{2}$Department of Astronomy, New Mexico State University, Las Cruces, New Mexico, USA\\
$^{3}$ Nevada Center for Astrophysics, University of Nevada, Las Vegas, 4505 South Maryland Parkway, Las Vegas, NV 89154, USA \\
$^{4}$ Department of Physics and Astronomy, University of Nevada, Las Vegas, 4505 South Maryland Parkway, Las Vegas, NV 89154, USA \\
}

\date{Accepted XXX. Received YYY; in original form ZZZ}

\pubyear{\the\year{}}

\begin{document}
\label{firstpage}
\pagerange{\pageref{firstpage}--\pageref{lastpage}}
\maketitle

\begin{abstract}
We present evidence that the streaming instability (SI) can form planetesimals
from millimeter grains inside axisymmetric pressure bumps, provided the bump is
not too strongly reinforced. We conducted three-dimensional local shearing-box
simulations of millimeter grains in a Gaussian pressure bump of amplitude
$A = 0.6$, at a resolution of $160/H$. We varied only the Newtonian
reinforcement timescale $t_{\rm reinf}$, the timescale on which the imposed bump
is restored against particle back-reaction. Our reference run T1 uses
$t_{\rm reinf} = 1\,\Omega^{-1}$. Particles pile up in a narrow axisymmetric
band at the minimum-headwind point, cross the Roche density, and collapse
through direct gravitational instability (GI). In our weakly reinforced run T100
with $t_{\rm reinf} = 100\,\Omega^{-1}$, the outcome is different. Particles
drift through the bump roughly three times more slowly, and dense SI-like
filaments develop across a wide radial region well before the Roche density is
reached. These results are consistent with a residence-time criterion, which states
that for the SI 
to grow, $t_{\rm cross} > t_{\rm grow}$, where $t_{\rm cross}$ is the time a particle
spends inside the region where the ratio $Z/\Pi$ of the solid abundance
$Z = \Sigma_p/\Sigma_g$ to the headwind parameter $\Pi = \Delta v/c_{\rm s}$ is
high enough for strong clumping, and $t_{\rm grow}$ is the SI growth time. T1
violates this criterion, and T100 satisfies it. Our results suggest that the
reinforcement timescale matters for millimeter grains because it helps set the
crossing time, and therefore whether the SI has time to grow. We caution,
however, that our reinforcement scheme is an idealised numerical construction
rather than a physical bump-forming mechanism such as a planet, and a more
definitive test of this picture will require simulations with a more physically
motivated bump.
\end{abstract}




\section{Introduction}

A central open problem in planet formation is how aerodynamically coupled dust grains are converted into planetesimals, the $\sim$1--100~km bodies that provide the building blocks of planetary cores. Growth by direct sticking is unlikely to operate uninterrupted from micron-sized grains to kilometre-sized bodies. As particles grow, their collision velocities increase, and laboratory and numerical studies suggest that collisions at millimetre--centimetre sizes often lead to bouncing or
fragmentation rather than continued growth \citep{2008ARA&A..46...21B,2010A&A...513A..56G,2010A&A...513A..57Z,2013A&A...559A..62W,2018SSRv..214...52B,2023ApJ...951L..16A}.  In addition, the gas in a protoplanetary disc is partially pressure supported and therefore orbits at a slightly sub-Keplerian speed. Solid particles experience this velocity difference as a headwind, lose angular momentum through aerodynamic drag, and drift radially towards the star \citep{1977MNRAS.180...57W,1986Icar...67..375N}. These barriers motivate models in which solids are first concentrated collectively and then collapse under their own gravity, bypassing the need for incremental collisional growth across the full planetesimal-size range.

The leading mechanism for this collective concentration is the streaming instability \citep[SI;][]{2005ApJ...620..459Y, 2007ApJ...662..627J}. The relative drift between gas and solids drives the SI. In essence, a local enhancement in the particle density backreacts on the gas, reduces the local headwind, and therefore slows the radial drift of particles in that region, while allowing dust concentrations to grow to larger amplitudes. More formally, in the low dust-to-gas regime, the SI can be understood as an epicyclic resonant drag instability arising from the coupling of particle drift to inertial modes through aerodynamic drag \citep{2020MNRAS.498.1239S}. The saturated state of the SI consists of dense, azimuthally extended particle filaments \citep{2007Natur.448.1022J, 2007ApJ...662..627J, 2010ApJS..190..297B}. If the density inside such filaments exceeds the Roche density,
\begin{equation}
    \rho_{\rm Roche} = \frac{9\Omega^2}{4\pi G},
\end{equation}
where $\Omega$ is the local orbital frequency, particle self-gravity can overcome Keplerian shear and drive collapse into bound planetesimals \citep{2007Natur.448.1022J,2016ApJ...822...55S,2017ApJ...847L..12S,2019ApJ...883..192A,2019ApJ...885...69L}. This picture of planetesimal formation is also consistent with Solar System constraints; bodies formed via the SI produce objects of size consistent with the largest Solar System bodies \citep{2009Icar..204..558M,2016ApJ...822...55S} and with angular momentum and spin obliquities in agreement with trans-Neptunian binaries \citep{2019NatAs...3..808N}.

The SI is not equally effective in all regions of a disc. Its non-linear outcome depends sensitively on the dimensionless particle stopping time $\tau_{\rm s}=\Omega t_{\rm stop}$, the vertically integrated solid abundance $Z=\Sigma_{\rm p}/\Sigma_{\rm g}$, and the strength of the pressure-supported headwind. The latter is commonly written as
\begin{equation}
    \Pi \equiv \frac{\Delta v}{c_{\rm s}},
\end{equation}
where $\Delta v$ is the difference between the Keplerian velocity and the azimuthal gas velocity, and $c_{\rm s}$ is the sound speed. Numerical surveys show that strong clumping requires Z to exceed a critical value that rises steeply for small particles \citep{2015A&A...579A..43C, 2017A&A...606A..80Y, 2021ApJ...919..107L,2025ApJ...981..160L,2026ApJ..1000..156L}. They also show that clumping is favoured when the pressure gradient, and hence the headwind, is weak \citep{2010ApJ...722L.220B}. This dependence is often expressed through the ratio $Z/\Pi$. \citet{2018ApJ...860..140S} showed that the overall structure of particle filaments produced by the SI scales with $Z/\Pi$, making this ratio a useful quantity for comparing SI behaviour across different pressure-gradient environments. These results suggest that planetesimal formation by the SI should be most efficient in regions where solids are concentrated, and the local headwind is reduced (e.g., larger values of $Z/\Pi$).
Observations of protoplanetary discs suggest that pressure bumps are common in
planet-forming environments. High-resolution ALMA observations, beginning with
the striking annular structure of HL~Tau, have shown that rings and gaps are
widespread and occur over a broad range of radii, widths, and amplitudes
\citep{2015ApJ...808L...3A, 2018ApJ...869L..41A, 2018ApJ...869L..42H}. Gas
kinematics and ring-width measurements provide further evidence for local
pressure-gradient perturbations, and for gas structures broader than the dust
rings they host \citep{2018ApJ...860L..12T, 2020MNRAS.495..173R}. Although
continuum rings alone do not prove ongoing planetesimal formation, their
ubiquity makes pressure bumps a compelling setting in which to study the SI.

Pressure bumps are promising sites for planetesimal formation because they act on both key ingredients of the SI. First, they collect drifting solids and raise the local value of $Z$. Secondly, they flatten the radial pressure gradient and reduce the headwind parameter $\Pi$. In a true pressure maximum, the drift velocity changes sign across the bump, producing a particle trap. At the exact pressure maximum, however, $\Pi=0$ and the relative drift that drives the SI vanishes, so planetesimal formation at the very centre of the trap requires a mechanism other than the SI. One such mechanism is direct gravitational instability: if a bump collects solids into a sufficiently narrow and dense band, that band can exceed the Roche density and fragment under its own gravity without the SI organising it first. We refer to this route as direct GI throughout. Even in a weaker pressure perturbation without a formal trap, the reduced headwind can substantially slow radial drift and move the system towards the SI-active regime. Thus, pressure bumps are not merely reservoirs of pebbles, but potentially privileged locations where the conditions for SI-driven planetesimal formation are naturally assembled.



This expectation has been tested directly in numerical simulations.
\citet{2021AJ....161...96C} carried out the first fully three-dimensional
self-gravitating simulations of the SI in a domain large enough to contain an
axisymmetric pressure bump comparable in scale to observed dust rings. They found that centimetre-sized particles can form planetesimals in modest bumps, even when the bump does not contain a pressure maximum and therefore does not halt radial drift completely. \citet{2022ApJ...927...52C} then explored how this outcome depends on the strength of the external forcing used to maintain the bump. In both sets of simulations, the bump was imposed through a Newtonian relaxation scheme that restores the gas density and azimuthal velocity towards a prescribed target profile on a reinforcement timescale $t_{\rm reinf}$. For centimetre-sized particles, planetesimal formation was remarkably resilient: increasing $t_{\rm reinf}$ by two orders of magnitude left the formation locations and the number of clumps largely unchanged \citep[see also][]{2021AJ....161...96C}.

The situation is less clear for millimetre-sized particles. This regime is
important because millimetre continuum observations are most directly sensitive to
mm-sized grains, and therefore the solids seen in ALMA rings are not necessarily
the centimetre-sized pebbles that form planetesimals most readily in simulations.
\citet{2022ApJ...933L..10C} performed a high-resolution study of mm-sized grains in
an axisymmetric pressure bump. Their simulation developed a long-lived particle
pile-up in which the local value of $Z/\Pi$ exceeded published clumping thresholds,
yet no strong SI-like clumping occurred. A companion run with a larger bump did
form planetesimals, but through direct GI rather than the SI: the bump concentrated
solids into a band that crossed the Roche density and fragmented, with no
filamentary structure beforehand. They argued this was due to the finite residence
time of particles inside the SI-favourable region. In a spatially localized bump,
particles do not experience the high-$Z/\Pi$ conditions indefinitely; instead, they
drift through the relevant region on a crossing time
\begin{equation}
    t_{\rm cross} \sim \frac{\ell}{|v_r|},
\end{equation}
where $\ell$ is the radial width of the high-$Z/\Pi$ region and $v_r$ is the
particle radial drift velocity. If $t_{\rm cross}$ is shorter than the SI growth
time $t_{\rm grow}$, the instability may fail to reach the non-linear filamentary
state even when the instantaneous local values of $Z$, $\Pi$, and $\tau_{\rm s}$
appear favourable. This led \citet{2022ApJ...933L..10C} to propose
$t_{\rm cross}>t_{\rm grow}$ as an additional criterion for SI-driven planetesimal
formation in finite-width pressure bumps.

This residence-time criterion places strong constraints on the role of
axisymmetric pressure bumps in forming planetesimals from mm-sized grains. However, the crossing time is not an externally fixed property of the bump: it depends directly on the particle drift speed. This is especially important in imposed-bump simulations, where the force that maintains the gas pressure structure will also modify the gas velocity field. Since solids are coupled to the gas by aerodynamic drag, an externally imposed torque on the gas can be communicated to the particles. Indeed, \citet{2022ApJ...927...52C} showed that stronger reinforcement produces faster particle drift. In their centimetre-grain simulations, particles in bumps reinforced on $t_{\rm reinf}=1\,\Omega^{-1}$ drifted approximately twice as fast as particles in otherwise similar bumps reinforced on $t_{\rm reinf}=10\,\Omega^{-1}$. In the mm-grain calculation of \citet{2022ApJ...933L..10C}, the measured drift speed through the high-$Z/\Pi$ region was substantially larger than the standard drift estimate, reducing the residence time available for SI growth.

This motivates a re-examination of mm-grain planetesimal formation in pressure
bumps. If the failure of strong clumping in \citet{2022ApJ...933L..10C} was
controlled by the short residence time of particles in the high-$Z/\Pi$ region, then changing the drift speed may also change the outcome. In particular, a strongly reinforced imposed bump may drive particles through the SI-favourable region faster than they would move in a more weakly reinforced pressure structure. The relevant question is therefore not only whether a pressure bump can produce favourable local values of $Z$, $\Pi$, and $\tau_{\rm s}$, but whether particles remain in that region long enough for the SI to grow.

In this paper we test this idea with controlled local simulations of mm-sized
particles in an axisymmetric pressure bump. We adopt the same particle size, solid
abundance, bump width, and numerical framework as the mm-grain pressure-bump
calculations of \citet{2022ApJ...933L..10C}, and focus on the case with a bump amplitude of 60\% relative to the background pressure ($A = 0.6$). We compare a strongly reinforced bump with $t_{\rm reinf}=1\,\Omega^{-1}$ to a weakly reinforced bump with $t_{\rm reinf}=100\,\Omega^{-1}$. Because the reinforcement timescale is the only physical parameter we change, differences between the runs can be attributed to the strength of the bump-maintaining torque. We examine its effect on the particle drift speed, the midplane kinetic energy, and the development of filamentary structure. We find that weaker reinforcement reduces the radial drift speed, allowing dense SI-like filaments to develop before the Roche density is reached. The midplane kinetic energy is also lower in the weakly reinforced run. The residence-time criterion proposed by \citet{2022ApJ...933L..10C} therefore remains the appropriate organizing principle, but the crossing time itself depends on how the bump is maintained.

The paper is organised as follows. In \autoref{sec:method} we describe the numerical methods, pressure-bump model, and simulation setup. In \autoref{sec:res} we present the evolution of the particle layer, compare the morphology of the strongly and weakly reinforced runs, and examine whether clumping is driven by the SI or by direct GI. In \autoref{sec:dis} we discuss the effect of reinforcement on particle drift and the implications for interpreting imposed-bump simulations. We summarize our conclusions in \autoref{sec:conclusions}.

\section{Numerical methods and Initial conditions}
\label{sec:method}
\subsection{Numerical method}
We use the same numerical methods as \citet{2022ApJ...933L..10C}, 
which we summarize briefly here. We perform three-dimensional local shearing-box simulations using the \textsc{athena} code in its purely hydrodynamic configuration. The gas is treated as a compressible, 
isothermal fluid with equation of state
\begin{equation}
    P = c_{\rm s}^{2}\rho_{\rm g},
\end{equation}
where $c_{\rm s}$ is the isothermal sound speed and $\rho_{\rm g}$ is the gas density. The solid component is represented by Lagrangian superparticles, which are coupled to the gas through aerodynamic drag. The backreaction of the particles on the gas is included. Particle self-gravity is calculated using a particle--mesh method. The particle mass is mapped onto the computational grid using a triangular-shaped cloud interpolation scheme, and the gravitational potential is obtained by solving the Poisson equation with a fast Fourier transform. The resulting gravitational acceleration is then interpolated back to the particle positions \citep{2016ApJ...822...55S}. 

The gas and particle variables use shearing-periodic boundary conditions 
in the radial direction and periodic boundary conditions in the 
azimuthal direction. In the vertical direction, we use the modified 
outflow boundary condition adopted by \citet{2022ApJ...933L..10C} and described in \citet{2018ApJ...862...14L}  in which the gas density is extrapolated into the ghost zones using an exponential profile. The gas density is renormalised at each timestep to maintain a constant total gas mass. The gravitational potential uses the same radial and azimuthal boundary conditions, together with open vertical boundaries.

We include neither magnetic fields nor externally driven turbulence. 
Consequently, any velocity fluctuations or turbulent motions that 
develop arise self-consistently from the gas--particle interaction and 
the source terms used to maintain the pressure bump. Further details of the numerical implementation are provided by \citet{2022ApJ...933L..10C} and the references therein.
\subsection{Initial conditions}
The shearing box is centered at a radius $r_0=50~{\rm au}$ in a
disk surrounding a solar-mass star. We adopt a gas surface-density profile that scales as $r^{-1}$,

\begin{equation}
\Sigma_g(r) =
\frac{M_{\rm disk}}{2\pi r_c r},
\label{eq:surface_density}
\end{equation}
where the total disk mass is $M_{\rm disk} = 0.09 M_{\odot}$ and the tapering
radius is $r_{c}=200~{\rm au}$. The central star's mass is $1 M_{\odot}$. This
gives a disk aspect ratio of
\begin{equation}
    \frac{H}{r}= \frac{c_{\rm s}}{v_{\rm K}}= 0.033 \left(\frac{r}{1~{\rm au}}\right)^{1/4},
    \label{eq:aspect_ratio}
\end{equation}
where $H=c_{\rm s}/\Omega$ is the gas scale height, $c_{\rm s}$ is the
isothermal sound speed, and $v_{\rm K}=r\Omega$ is the Keplerian
velocity. The temperature profile of the disk is given by the following equation
\begin{equation}
    T(r)= 280\left(\frac{r}{1~{{\rm au}}}\right)^{-1/2} {\rm K}.
    \label{eq:temperature_profile}
\end{equation}
Before the pressure bump is introduced, the gas is initialised in hydrostatic equilibrium
\begin{equation}
    \rho_{\rm g}(z)=\rho_{0}\exp\left(-\frac{z^{2}}{2H^{2}}\right),
    \label{eq:initial_gas_density}
\end{equation}
where $\rho_{0}$ is the initial midplane gas density. The background gas density
is uniform in the radial and azimuthal directions within the local domain. The
particles have a solid-to-gas ratio of $Z=0.01$ where
\begin{equation}
    Z\equiv\frac{\Sigma_{\rm p}}{\Sigma_{\rm g}}=0.01.
    \label{eq:initial_metallicity}
\end{equation}
The horizontal distribution of the particles is initially uniform, while their
vertical positions are drawn from a Gaussian distribution with a scale height of
$H_{\rm p}=0.025H$. Perturbations are added to the particle positions to
seed the growth of particle--gas instabilities; no perturbations are applied to
the gas. Similar to \citet{2022ApJ...933L..10C}, we adopt a single particle size
of $a = 1~{\rm mm}$. The dimensionless stopping time is
\begin{equation}
    \tau_{\rm s}=\Omega t_{\rm stop} = \sqrt{\frac{\pi}{8}}\,
    \frac{\rho_{\rm s} a}{\rho_{\rm g} c_{\rm s}}\,\Omega,
    \label{eq:stokes}
\end{equation}
where $\rho_{\rm s}$ is the internal density of the solid material. We hold the
particle size fixed at $a = 1~{\rm mm}$ and compute $t_{\rm stop}$ from the local
gas density, so $\tau_{\rm s}$ is not constant across the domain: because
$\tau_{\rm s}\propto1/\rho_{\rm g}$, the stopping time decreases inside the
pressure bump, where the gas density is enhanced. Particles begin with
$\tau_{\rm s}\simeq0.0123$, and this value varies as they drift through the bump.
The global radial pressure gradient is retained through the dimensionless headwind
parameter
\begin{equation}
    \Pi\equiv \frac{\Delta v}{c_{\rm s}} = -\frac{1}{2} \frac{c_{\rm s}}{v_{\rm K}} \frac{{\rm d}\ln P}{{\rm d}\ln r} = 0.12,
    \label{eq:background_pressure}
\end{equation}
where $\Delta v$ is the difference between the Keplerian particle
velocity and the partially pressure-supported azimuthal gas velocity. This way, particles
experience inward radial drift even before we account for the local modification
of the pressure gradient produced by the pressure bump. Finally, the strength of
the particle self-gravity is computed using the following equation
\begin{equation}
    \tilde{G} \equiv \frac{4\pi G \rho_0}{\Omega^2} = 0.2,
    \label{eq:self_gravity_strength}
\end{equation}
\autoref{eq:self_gravity_strength} leads to a Toomre (\citealt{1964ApJ...139.1217T})
$Q$ value of $Q \approx 8$, so the gas disk is gravitationally stable.

In this study, we impose the same axisymmetric pressure bump profile as
\citet{2022ApJ...933L..10C}. The gas density consists of a Gaussian radial
enhancement which is superimposed on the vertically stratified background disk
\begin{equation}
    \widehat{\rho}_{\rm g}(x,z) = \rho_0 \left[ 1+A\exp\left(-\frac{x^2}{2w^2}\right) \right] \exp\left(-\frac{z^2}{2H^2}\right),
    \label{eq:target_bump_density}
\end{equation}
where $A$ and $w$ are the amplitude and the radial width of the bump,
respectively. Across all the simulations, we adopt $A = 0.6$ and $w=1.14H$. The
target azimuthal gas velocity, $\widehat{u}_y(x)$, is calculated such that the
imposed density profile is in a radial geostrophic balance. We maintain the gas
density and the azimuthal velocity using a Newtonian relaxation. At every
timestep, it is adjusted according to the following equations
\begin{equation}
    \Delta\rho_{g} = -\left( \rho_{\rm g}-\widehat{\rho}_{g}\right)
    \frac{\Delta t}{t_{\rm reinf}},
    \label{eq:density_reinforcement}
\end{equation}

and

\begin{equation}
    \Delta u_y = -\left( u_y-\widehat{u}_y \right) \frac{\Delta t}{t_{\rm reinf}},
    \label{eq:velocity_reinforcement}
\end{equation}
where $t_{\rm reinf}$ is the pressure-bump reinforcement time. This
parameter determines how rapidly deviations from the target density
and velocity profiles are removed.

\subsection{Simulation setup}
\label{sec:setup}
Our experiment consists primarily of two simulations: T1 and T100.
We perform the simulations in a local Cartesian shearing box, where
$x$, $y$, and $z$ correspond to the radial, azimuthal, and vertical
directions of the disk, respectively. All three runs use the same disk
model, particle properties, self-gravity strength, and Gaussian
pressure bump. The runs use the same radial and azimuthal domain sizes
$L_x \times L_y = 9H \times 0.2H$. We choose a long radial domain so that the box
can contain the entire radial extent of the pressure bump. This allows us to
follow particles as they drift toward the pressure bump, pass the minimum
headwind region, and continue downstream. The simulation parameters are
summarised in \autoref{tab:sim}.

\begin{table*}
\centering
\caption{Summary of the simulation parameters. Runs T1, T100, and T100-HR share
the same disk model, particle size, solid abundance, and pressure-bump profile,
and differ only in the reinforcement timescale $t_{\rm reinf}$, the vertical
domain size required to converge at that timescale, and the grid resolution. Run
NoBump is a reference SI simulation without a bump.}
\label{tab:sim}
\scriptsize
\setlength{\tabcolsep}{18pt}
\renewcommand{\arraystretch}{1.25}
\begin{tabular}{lcccccccc}
\hline
\hline
Name &
$\tau_{\rm s}$ &
$Z$ &
$A$ &
$w$ &
$L_x \times L_y \times L_z$ &
Resolution &
$t_{\rm reinf}$ \\
\hline
Run T1 & $0.0123$ & $0.01$ & $0.6$ & $1.14H$ &
$9H \times 0.2H \times 0.8H$ & $160/H$ & $1\Omega^{-1}$ \\

Run T100 & $0.0123$ & $0.01$ & $0.6$ & $1.14H$ &
$9H \times 0.2H \times 3.2H$ & $160/H$ & $100\Omega^{-1}$ \\

Run T100-HR & $0.0123$ & $0.01$ & $0.6$ & $1.14H$ &
$9H \times 0.2H \times 3.2H$ & $320/H$ & $100\Omega^{-1}$ \\

Run NoBump & $0.0097$ & $0.03$ & --- & --- &
$0.4H \times 0.2H \times 3.2H$ & $640/H$ & --- \\
\hline
\hline
\end{tabular}
\end{table*}

\textbf{Run T1:} This run is identical to the run described in
\citet{2022ApJ...933L..10C} where $A = 0.6$ and $w = 1.14H$. This run has a
vertical domain size of $L_z=0.8H$, giving a full domain of
$L_x \times L_y \times L_z= 9H \times 0.2H \times 0.8H$. We reinforce the
pressure bump on a timescale $t_{\rm reinf}=1\Omega^{-1}$. The run uses a
resolution of $160/H$, which equates to $1440 \times 32 \times 128$ spatial grid
cells.

\textbf{Run T100:} We change the reinforcement time from $1\Omega^{-1}$ to
$100\Omega^{-1}$ to examine how a more weakly maintained pressure bump affects
the coupled gas--particle evolution. This run also uses a taller vertical domain,
$L_x \times L_y \times L_z= 9H \times 0.2H \times 3.2H$. The taller box is not an
independent choice but a requirement of the longer reinforcement time.
\citet{2022ApJ...927...52C} showed that for weakly reinforced bumps, tall domains lead to the pressure bump maintaining its initial profile: particles are in contact with the gas only
near the midplane, so in a short box the midplane gas dynamics are dominated by
the particles and the bump profile is distorted away from its target shape. A
taller box contains enough gas above and below the midplane to refill the deficit
produced by particle back-reaction. This is only an issue at large
$t_{\rm reinf}$, because for short reinforcement times the midplane gas is reset
towards the target profile on every timestep. Retaining $L_z = 0.8H$ at
$t_{\rm reinf}=100\Omega^{-1}$ would therefore not have produced the same bump.
The run uses the same resolution per $H$ as Run T1, equating to
$1440 \times 32 \times 512$ spatial grid cells.

\textbf{Run T100-HR:} This run is identical to T100 in every respect except that
we double the resolution to $320/H$, giving a full grid of
$2880 \times 64 \times 1024$ cells. The domain, physical parameters, and
reinforcement timescale are unchanged. Its purpose is to test whether the
SI-driven filament structure that develops in T100 is a genuine feature of the
pressure bump or a numerical artefact of the $160/H$ resolution. At
$\tau_{\rm s} \simeq 0.0123$, the fastest-growing SI modes are marginally
resolved at $160/H$ \citep{2017A&A...606A..80Y}, so doubling the resolution
provides a direct test of convergence for the T100 result.

\textbf{Run NoBump:} This is a standard SI reference simulation with no pressure
bump. We use it to compare the morphology of the filaments that develop inside
the T100 bump against those of a well-resolved SI simulation without a pressure
bump. The domain is $L_x \times L_y \times L_z = 0.4H \times 0.2H \times 3.2H$ at
a resolution of $640/H$, giving a full grid of $256 \times 128 \times 2048$
cells. The radial extent is chosen to match the width of the SI-active region
inside the T100 bump, and the vertical extent matches T100 so that vertical
stratification is treated identically in both runs. We adopt $Z = 0.03$ and
$\Pi = 0.05$, placing the run in the standard strong-clumping regime probed by
\citet{2017A&A...606A..80Y}, \citet{2021ApJ...919..107L},
\citet{2025ApJ...981..160L}, and \citet{2026ApJ..1000..156L}. The purpose of
this run is therefore to provide a reference for the morphology of nonlinear SI
filaments rather than a parameter-matched reproduction of the SI-active region
in T100. All other physical parameters, boundary conditions, and numerical
methods match those described in \autoref{sec:method}.

\section{Results}
\label{sec:res}
Here we present the results of our simulations. We first give an
overview of the morphological evolution of the particle layer in
T1, T100, and T100-HR.

\subsection{Evolution of the particle layer}
\label{sec:SI}
\begin{figure}
	\includegraphics[width=\columnwidth]{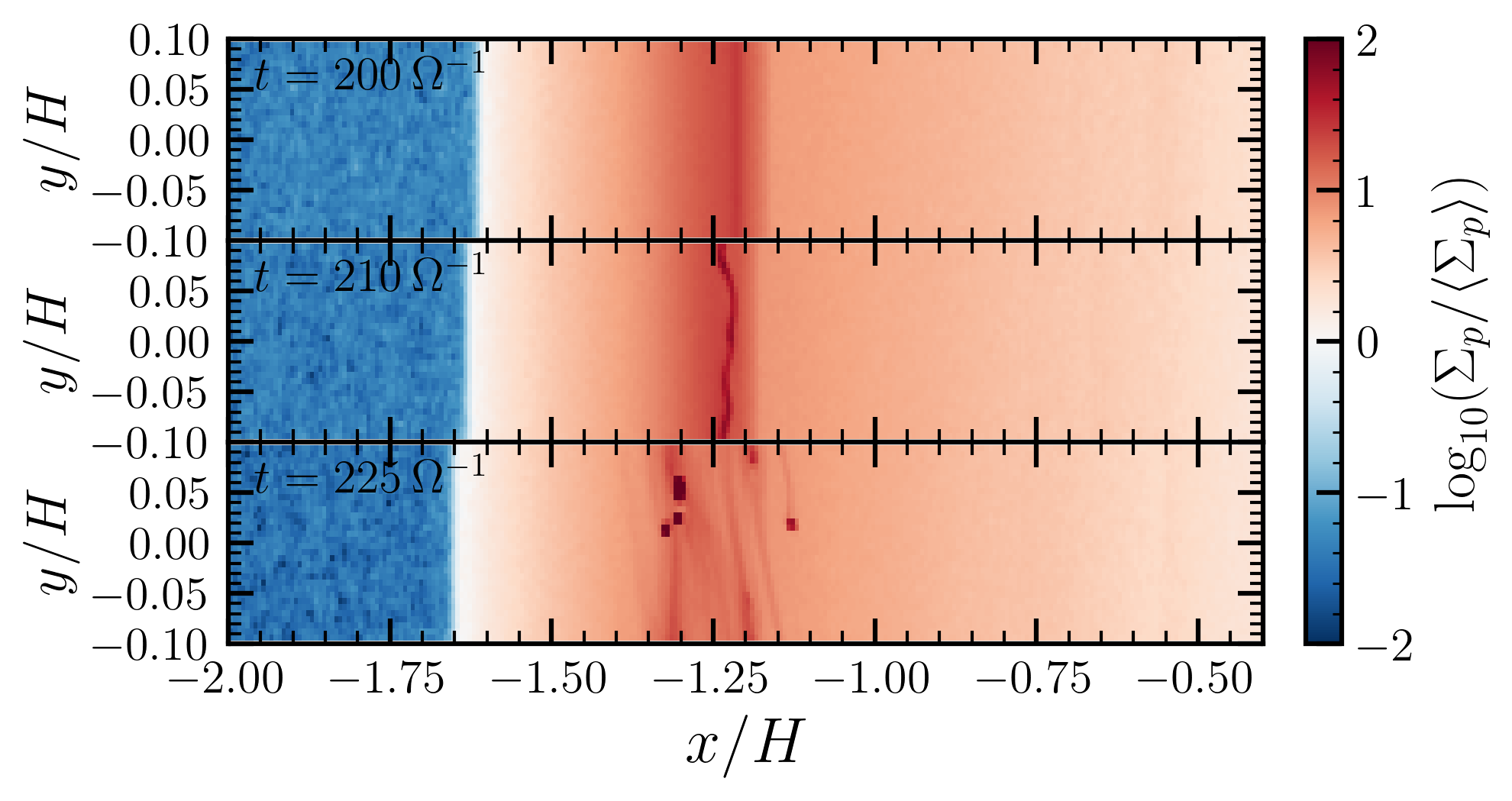}
  \caption{Particle surface density $\Sigma_{\rm p}$ in run T1 at $t = 200$,
  $215$, and $225\,\Omega^{-1}$, normalised by the initial box-averaged value
  $\langle \Sigma_{\rm p} \rangle$. The transition to nonaxisymmetric structure
  coincides with the particle density crossing the Roche threshold, and is a
  signature of direct gravitational instability.}
    \label{fig:eT1_surf}
\end{figure}

\textbf{Run T1:} This run is identical to the low-res, $A=0.6$ run of
\citet{2022ApJ...933L..10C}. This is the reference case in this study for
comparison purposes with the weakly reinforced simulations.
\autoref{fig:eT1_surf} shows three snapshots of the simulations at $t=200$,
$215$, and $225\,\Omega^{-1}$. At $t=200\,\Omega^{-1}$, there is a nearly
axisymmetric narrow band of particle pileup near the point of minimum headwind at
$x\simeq -1.2H$. By $t=215\,\Omega^{-1}$, the particle band has developed
nonaxisymmetric structure, and by $t=225\,\Omega^{-1}$ it has broken into compact
clumps. As shown by \citet{2022ApJ...933L..10C}, the appearance of
nonaxisymmetric structure in this run coincides with the particle density
crossing the Roche density. This is the signature of direct gravitational
instability: the pressure bump concentrates particles into a narrow band until
the band becomes self-gravitating and fragments into gravitationally bound
objects.

\begin{figure}
	\includegraphics[width=\columnwidth]{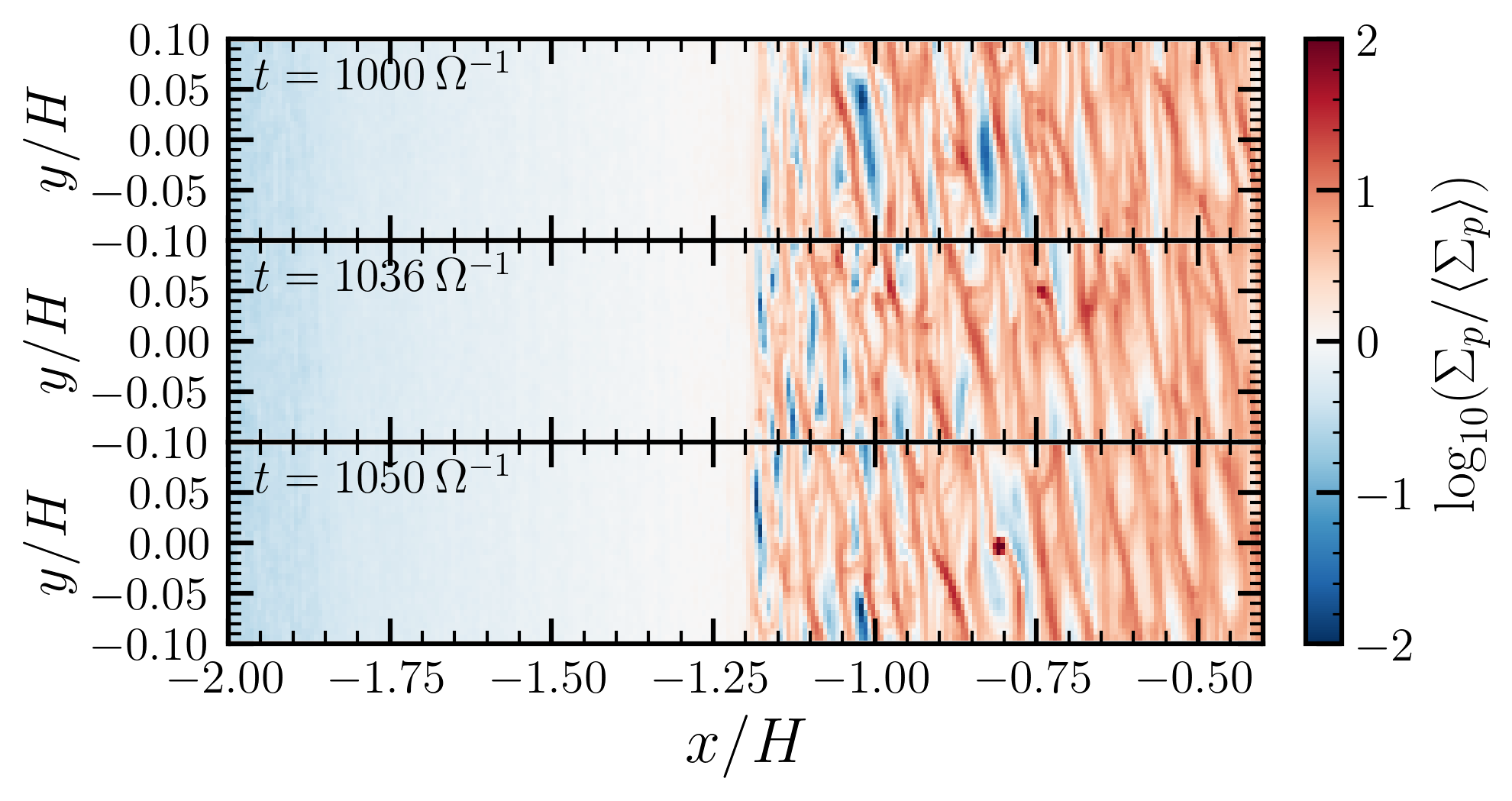}
  \caption{Particle surface density $\Sigma_{\rm p}$ in run T100 at $t = 1000$,
  $1036$, and $1050\,\Omega^{-1}$, normalised by the initial box-averaged value
  $\langle \Sigma_{\rm p} \rangle$. At $t = 1000\,\Omega^{-1}$, particles are
  already organised into narrow azimuthally extended filaments that span
  $x/H \simeq -1.2$ to $-0.5$, rather than a single band at the minimum-headwind
  point as seen in run T1.}
    \label{fig:t100_surf}
\end{figure}

\textbf{Run T100:} \autoref{fig:t100_surf} shows three snapshots of
T100 at $t = 1000$, $1036$, and $1050\,\Omega^{-1}$. At
$t = 1000\,\Omega^{-1}$, the particle layer is already organised into narrow
azimuthal filaments that extend across the bump-feeding region from
$x/H \simeq -1.2$ to $-0.5$. Rather than a single narrow band at the
minimum-headwind point, the filaments occupy a radially extended region spanning
roughly $0.7H$, from the minimum-headwind point at $x/H \simeq -1.2$ outward to
$x/H \simeq -0.5$. By $t = 1036\,\Omega^{-1}$ the filaments have drifted and
sharpened, and a compact overdensity has begun to form at $x/H \simeq -0.8$. By
$t = 1050\,\Omega^{-1}$ this overdensity has become a gravitationally bound
clump. Unlike T1, where nonaxisymmetric structure appeared only once the particle
density had crossed the Roche threshold, in T100 the filaments are present while
the particle density is still climbing well below Roche. The particle layer is
organised aerodynamically first, and only later does the densest filament reach
the density required for gravitational collapse.

\textbf{Run T100-HR:} The higher-resolution version of T100 reproduces the same
morphological evolution. The particle layer organises into narrow azimuthal
filaments across the same radial region as in T100. Within the densest filament,
a compact overdensity forms and eventually becomes gravitationally bound at the
same radial location. The sequence is the same as in T100: the particle layer is
organised aerodynamically first, and only later does the densest filament reach
the density required for gravitational collapse. The filaments emerge earlier in
T100-HR than in T100, consistent with the finding of
\citet{2017A&A...606A..80Y} that better-resolved SI modes grow faster. Doubling
the resolution therefore does not change the qualitative outcome of the run.

\subsection{Distinguishing SI-driven clumping from direct GI}
\label{sec:SI not GI}

As shown in \autoref{sec:SI}, the weakly reinforced runs do not follow
the same route to planetesimal formation as T1. In T1, particles collect
into a single narrow band that later breaks up through direct
gravitational instability. In contrast, T100 and T100-HR develop
extended filamentary structures in which particles concentrate before
collapse. In this section, we show that this difference is also
reflected in the time evolution of the maximum particle density.

\begin{figure}
	\includegraphics[width=\columnwidth]{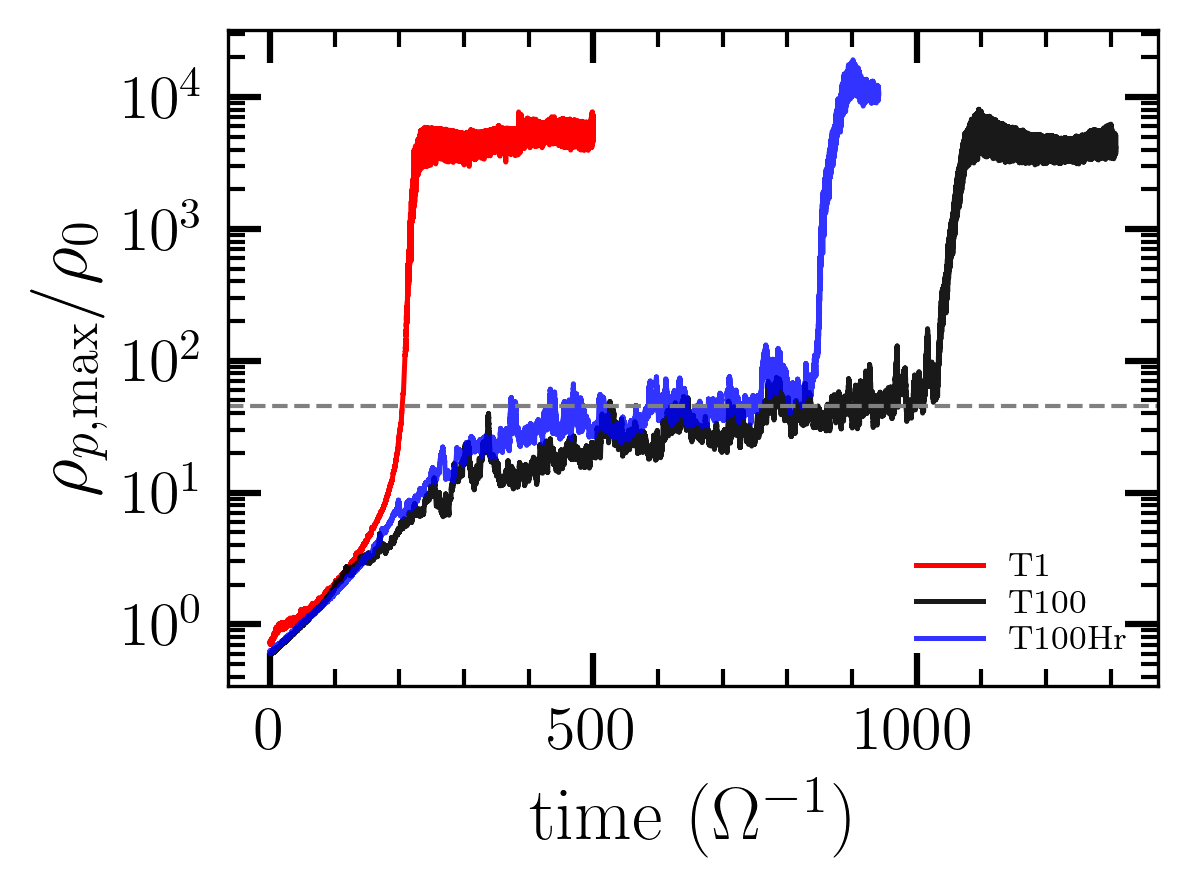}
    \caption{
    Maximum particle density as a function of time for T1, T100, and
    T100-HR. The dashed horizontal line marks the Roche density. In T1, the
    rapid rise in $\rho_{\rm p,max}$ reaches the Roche density at
    $t\simeq200\,\Omega^{-1}$, coincident with the breakup of the
    particle band shown in \autoref{fig:eT1_surf}. The weakly reinforced runs
    evolve for several hundred orbits in a pre-collapse state before the final
    runaway increase in particle density. T100-HR follows the same sequence as
    T100 at twice the resolution.}
    \label{fig:roche}
\end{figure}

\autoref{fig:roche} shows $\rho_{\rm p,max}$ as a function of time for
T1, T100, and T100-HR. The dashed horizontal line marks the Roche
density used in this study. In T1, $\rho_{\rm p,max}$ rises rapidly and
crosses the Roche density at $t\simeq200\,\Omega^{-1}$. This coincides
with the breakup of the narrow particle band and the formation of
gravitationally bound objects through direct gravitational instability, as also
found by \citet{2022ApJ...933L..10C}.

In T100 and T100-HR, $\rho_{\rm p,max}$ grows more slowly. Both runs spend
several hundred orbits in a pre-collapse state. During this phase, the particles
concentrate into SI-like filaments rather than into a single dense band.

\begin{figure}
	\includegraphics[width=\columnwidth]{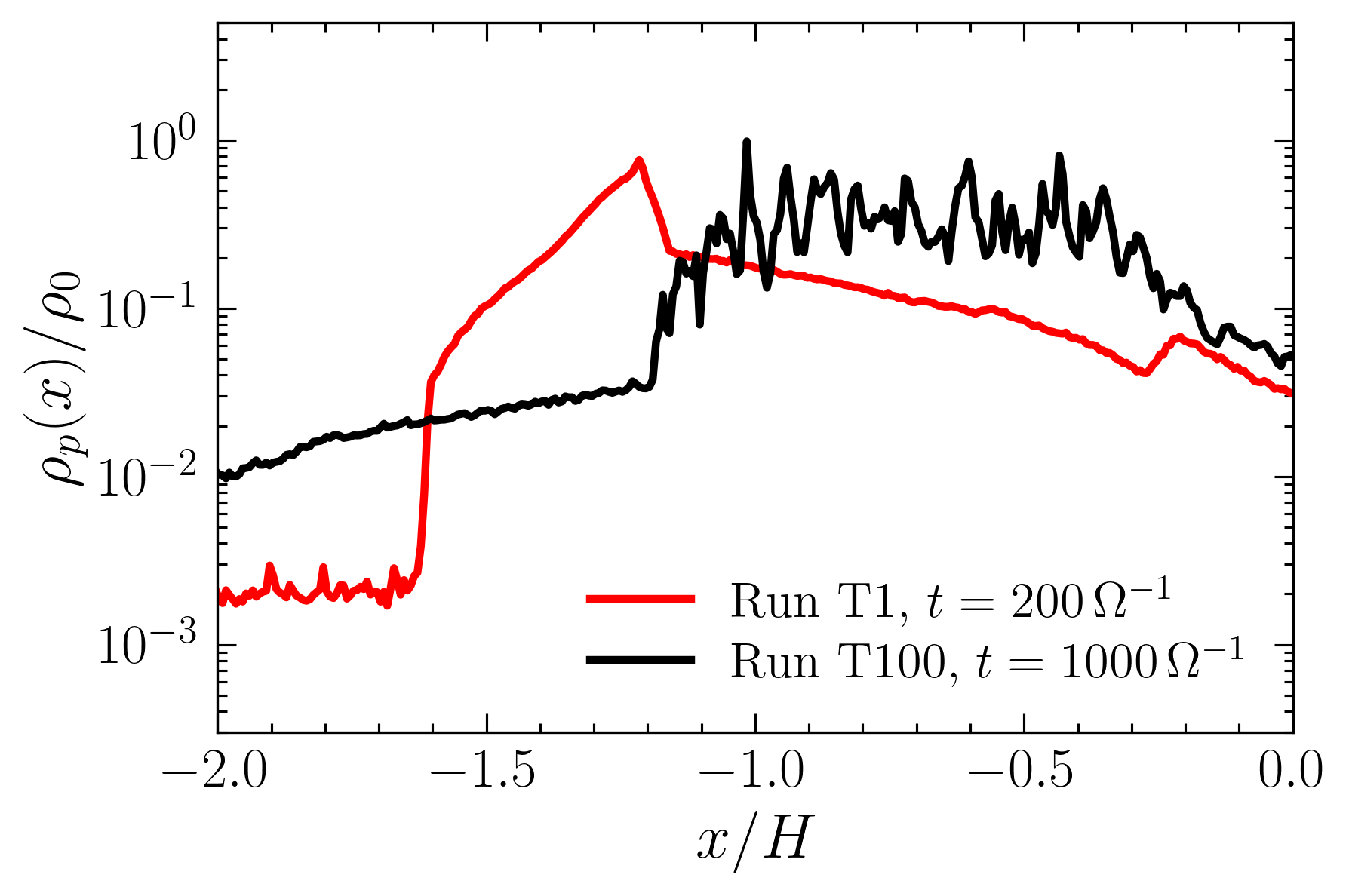}
    \caption{Radial profile of the particle volume density $\rho_{\rm p}(x)$,
    averaged over $y$ and $z$. Run T1 (red) is shown at $t = 200\,\Omega^{-1}$,
    at the onset of gravitationally bound object formation via direct
    gravitational instability. Run T100 (black) is shown at
    $t = 1000\,\Omega^{-1}$, during the pre-collapse phase when SI-driven
    filaments have become visible in the radial profile.}
    \label{fig:rhop_x}
\end{figure}

We now examine whether the particle concentrations form at the same
radial location in these runs. \autoref{fig:rhop_x} shows the radial
particle-density profile near the onset of clumping. In T1, the profile is
dominated by a single sharp peak near $x\simeq-1.25H$, corresponding to the
narrow band at the minimum-headwind point. T100 shows a qualitatively different
structure. Instead of an isolated peak at the minimum-headwind location, the
particle overdensities are spread across a wide radial region from
$x\simeq-1.2H$ to $\simeq-0.3H$, with several narrow maxima superimposed on the
underlying profile. These maxima are the SI filaments identified in
\autoref{sec:SI}. Together with the slower growth of $\rho_{\rm p,max}$ shown in
\autoref{fig:roche}, the radial profile shows that T100 reaches planetesimal
formation through a different sequence: SI-like filamentary concentrations form
first, spread across a wide region extending outward from the minimum-headwind
point, and only later does the densest filament collapse gravitationally.

\subsection{Local conditions for SI-like filament formation}
\label{sec:si_conditions}
\autoref{sec:SI not GI} showed that runs T1 and T100 reach collapse through different routes. We now ask whether the local conditions in T100 are consistent with SI-driven particle concentration, and how they differ from the conditions in T1. 

The presence of a pressure bump leads to radial variation in three quantities central to the streaming instability: the local dust loading, the local headwind, and the local stopping time. The stopping time varies through the bump because $\tau_{\rm s} \propto 1/\rho_{\rm g}$ and the gas density changes, but the change is modest (i.e., our bumps have relatively low amplitude). Thus, here, we only focus on how $Z$ and $\Pi$ vary; we measure the dust-to-gas ratio $Z$ and the headwind parameter $\Pi$ (defined in \autoref{sec:method}) as functions of radial position and time, $Z(x,t)$ and $\Pi(x,t)$, and calculate the ratio $Z/\Pi$ as it has been shown to be a key parameter for the structure of SI filaments \citep{2018ApJ...860..140S}.


\begin{figure}
	\includegraphics[width=\columnwidth]{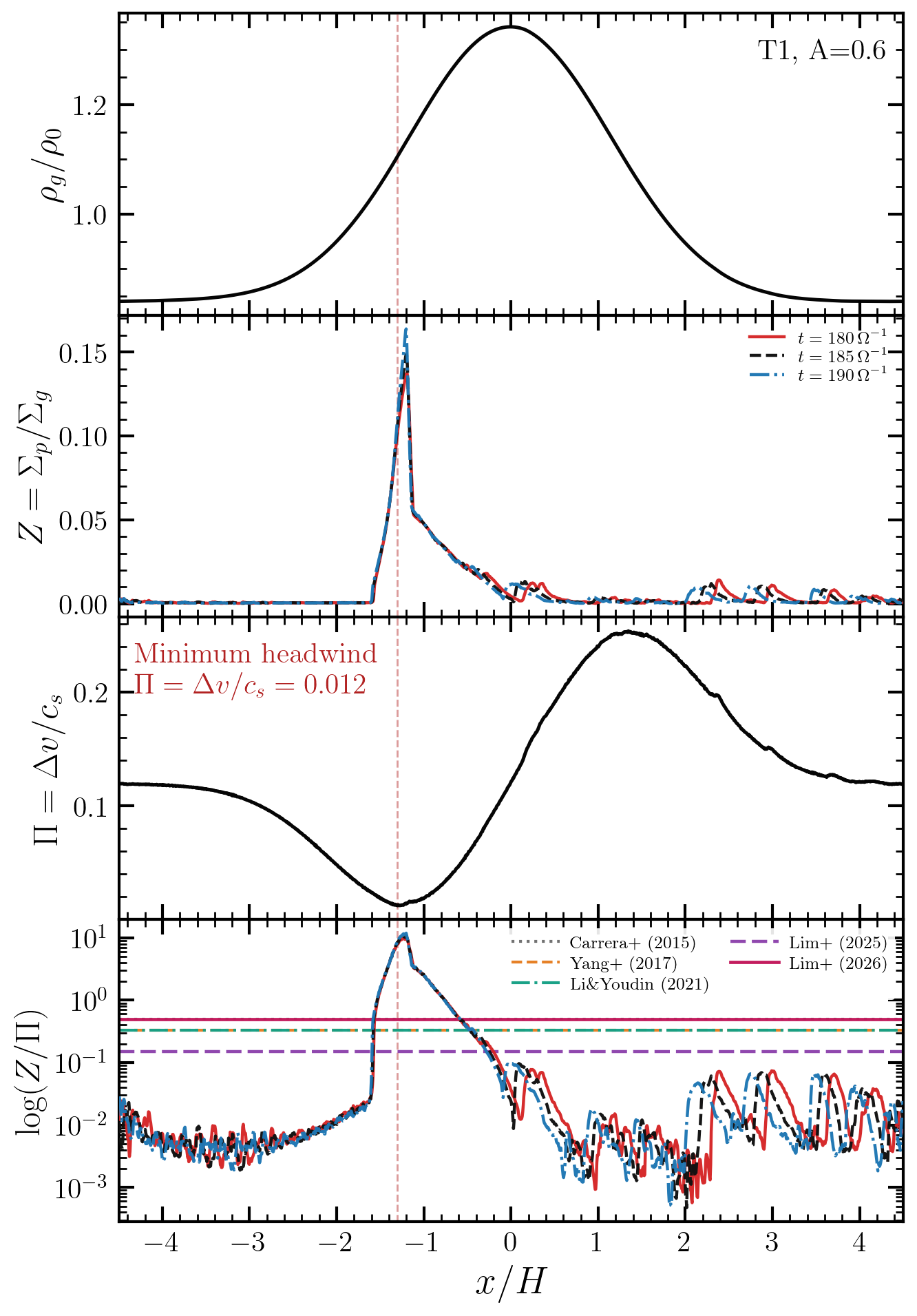}
    \caption{Radial profiles of run T1 near the onset of collapse, at $t = 180$, $185$, and $190\,\Omega^{-1}$. From top to bottom: the midplane gas density $\rho_g/\rho_0$, which sets the location of the bump; the dust-to-gas ratio $Z = \Sigma_p/\Sigma_g$; the local headwind parameter $\Pi = \Delta v/c_s$; and the ratio $Z/\Pi$. The horizontal lines in the bottom panel mark the strong-clumping thresholds of \citet{2015A&A...579A..43C}, \citet{2017A&A...606A..80Y}, \citet{2021ApJ...919..107L}, \citet{2025ApJ...981..160L}, and \citet{2026ApJ..1000..156L}.}
    \label{fig:zpi}
\end{figure}

\begin{figure}
    \includegraphics[width=\columnwidth]{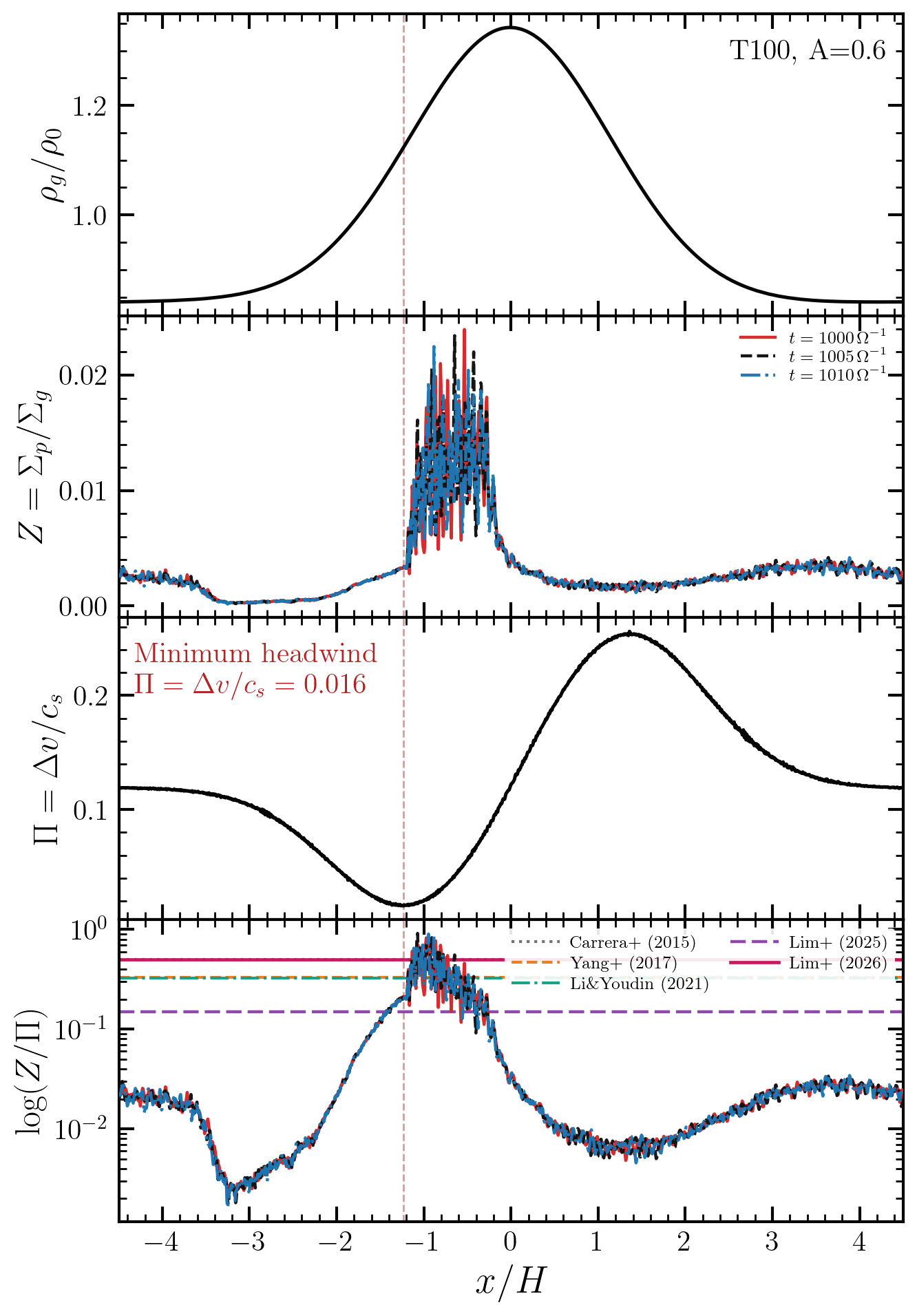}
    \caption{Radial profiles of run T100 during the pre-collapse phase, at $t = 1000$, $1005$, and $1010\,\Omega^{-1}$. From top to bottom: the midplane gas density $\rho_g/\rho_0$, which sets the location of the bump; the dust-to-gas ratio $Z = \Sigma_p/\Sigma_g$; the local headwind parameter $\Pi = \Delta v/c_s$; and the ratio $Z/\Pi$. The horizontal} lines in the bottom panel mark the strong-clumping thresholds of \citet{2015A&A...579A..43C}, \citet{2017A&A...606A..80Y}, \citet{2021ApJ...919..107L}, \citet{2025ApJ...981..160L}, and \citet{2026ApJ..1000..156L}.
    \label{fig:zpi_T100}
\end{figure}

\autoref{fig:zpi} and \autoref{fig:zpi_T100} compare these quantities for T1 near collapse
($t = 180$--$190\,\Omega^{-1}$) and for T100 in its pre-collapse phase ($t = 1000$--$1010\,\Omega^{-1}$). The two runs occupy strikingly different regions of parameter space. In T1, $Z(x)$ shows a single narrow spike at the minimum-headwind point with $Z_{\rm max} \simeq 0.15$ concentrated within a radial band no wider than $\sim 0.1\,H$. The combination of high loading and low $\Pi_{\rm min} = 0.012$ drives $Z/\Pi$ to values above 10, more than an order of magnitude above the strong-clumping thresholds of \citet{2015A&A...579A..43C}, \citet{2017A&A...606A..80Y}, \citet{2021ApJ...919..107L}, \citet{2025ApJ...981..160L}, and \citet{2026ApJ..1000..156L}.

T100 presents a different behaviour. The peak $Z$ is much lower
($\sim 0.022$) and the elevated-$Z$ region spans from $x/H \simeq -1.2$ to $\simeq -0.4$, roughly an order of magnitude wider than in T1. Within this region, $Z(x)$ is not smooth but shows multiple sub-peaks: these are the azimuthally extended filaments already visible in the surface density snapshots (\autoref{fig:t100_surf}). The corresponding $Z/\Pi$ profile just crosses the strong-clumping thresholds of \citet{2017A&A...606A..80Y}, \citet{2021ApJ...919..107L}, \citet{2025ApJ...981..160L}, and \citet{2026ApJ..1000..156L} across the same broad region, rather than spiking sharply at a single location. The comparison shows that reaching a large $Z/\Pi$ is not enough to determine whether or not the SI will emerge. T1 exceeds every published clumping threshold by a wide margin, but the particles pass through the high-$Z/\Pi$ region too quickly for the SI to organise the particle layer before gravitational instability sets in. The narrowness of the region likely contributes as well, since it sets the crossing time itself, but the primary constraint is temporal. T100 crosses the thresholds by a smaller margin, but keeps the particle layer inside the favourable region for long enough that SI-driven filaments develop before gravitational collapse. In this sense, the reinforcement time controls the role of the bump. In T1, the bump acts as a rapid collector, producing a narrow band that the particles cross before the SI can grow, so gravitational instability sets in directly.\footnote{It is worth noting that while the bump is not sufficiently strong to halt radial drift of particles, it {\it is} strong enough to slow down drift enough for direct GI to take over} In T100 the bump acts as a longer-lived reservoir in which particles remain long enough for the SI to grow before self-gravity takes over. This is consistent with the residence-time picture proposed by \citet{2022ApJ...933L..10C}: strong local conditions alone are insufficient; the particle layer must remain in the favourable region long enough for the instability to reach nonlinear amplitudes.

\begin{figure*}
    \centering
    \includegraphics[width=0.48\textwidth]{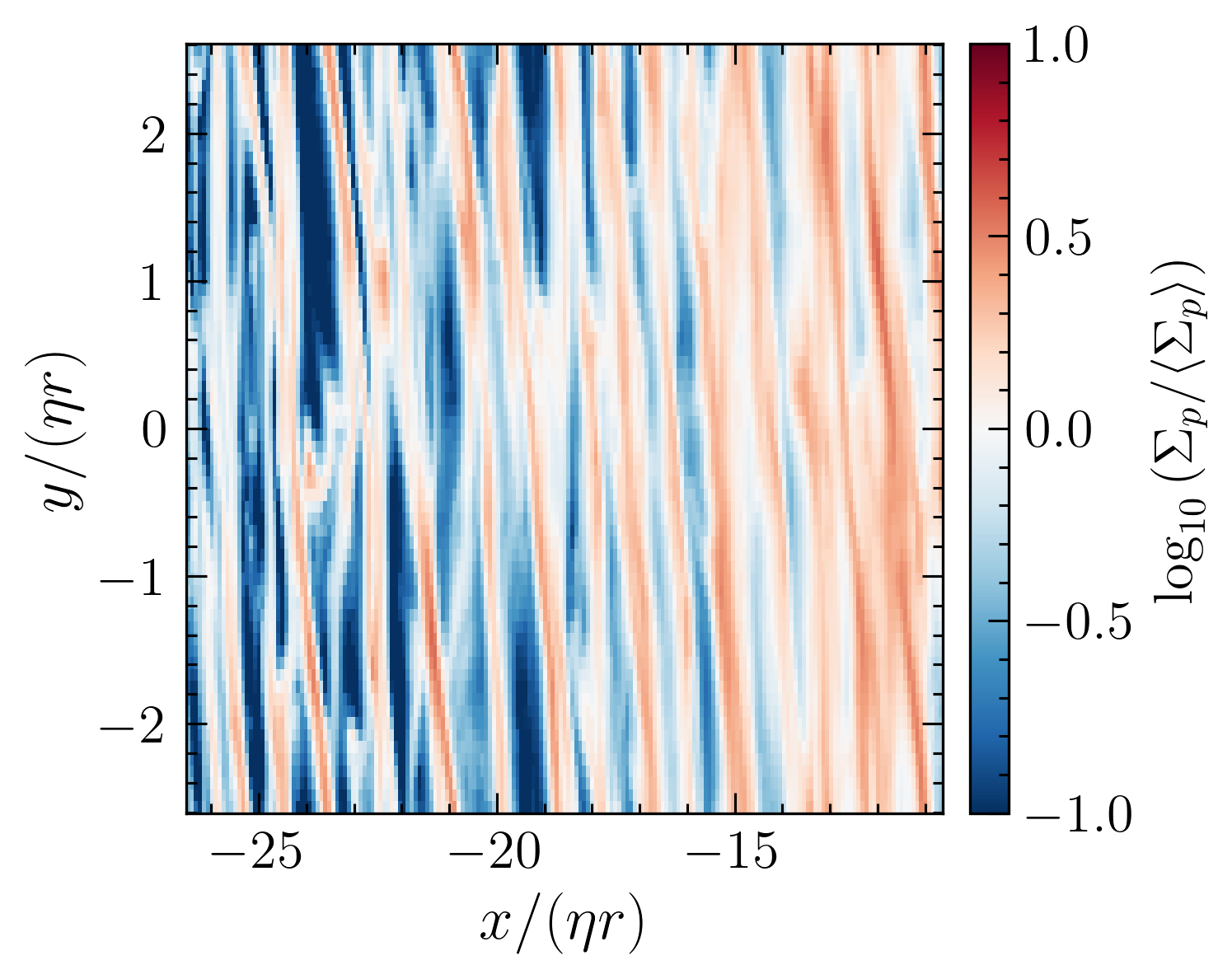}
    \includegraphics[width=0.48\textwidth]{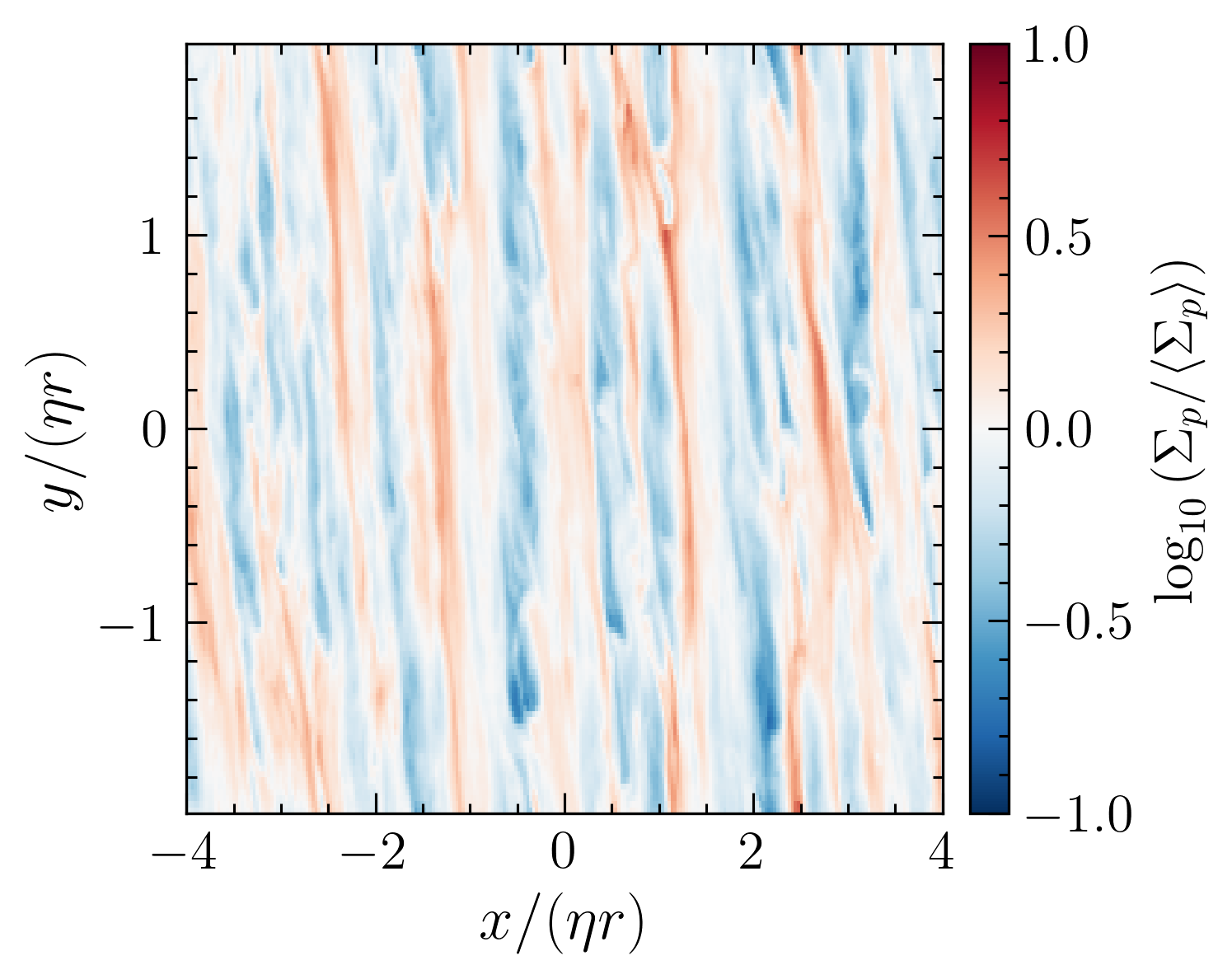}
  \caption{Particle surface density $\Sigma_p$ normalised by its
    box-averaged value $\langle \Sigma_p \rangle$, on a logarithmic scale. Left: run T100 at $160/H$ resolution, inside the pressure bump. Right: no-bump reference simulation at $640/H$ resolution. Both snapshots are taken during the nonlinear state. Radial and azimuthal coordinates are shown in units of $\eta r$, computed with the local headwind at the centre of the SI-active region ($\Pi \simeq 0.025$) for T100 and with the global $\Pi = 0.05$ for the no-bump run. Both runs display azimuthally elongated filaments with a characteristic spacing of about $\eta r$.}
    \label{fig:sigma_p}
\end{figure*}

We now take a closer look at the morphology of the particle concentrations within the non-linear SI state. Figure~\ref{fig:sigma_p} shows the particle surface density $\Sigma_p$ in run T100 alongside a high-resolution no-bump reference simulation at $640/H$ resolution. Both snapshots are taken at a similar stage in the nonlinear evolution. In each panel we normalise $\Sigma_p$ by its box-averaged value $\langle \Sigma_p \rangle$ and plot the radial and azimuthal coordinates in units of $\eta r$, the natural length scale for the SI. Because $\eta$ varies with position through the bump in T100, we compute $\eta r$ using the local headwind at the centre of the SI-active region ($\Pi \simeq 0.025$); for the no-bump reference simulation we use the global $\Pi = 0.05$. Both runs show qualitatively similar azimuthally elongated filamentary structure on scales of order $\eta r$. The reference run at four times the resolution shows the same morphology and the same spacing. This is consistent with the SI filament separation of order $\eta r$ reported by \citet{2017A&A...606A..80Y} and \citet{2021ApJ...919..107L}.


\subsection{Location of gravitationally bound object formation}
\label{sec:location}

An additional result of our comparison is that the gravitationally bound objects in T1 and T100 form at different radial locations. In T1, the compact overdensity that reaches the Roche density forms at $x/H \simeq -1.25$, at the minimum-headwind point where the pressure bump concentrates particles into a narrow band. In T100, the densest filament that reaches the Roche density forms at $x/H \simeq -0.8$, offset from the minimum-headwind point by roughly $0.4H$.

This offset reflects the different physical mechanisms operating in the two runs. In T1, the bump collects particles into a tight radial band, and the location of gravitational collapse is set by the point of minimum headwind, where the bump concentrates particles most efficiently. In T100, the SI develops filaments across the wide low-headwind region ($x/H \simeq -1.2$ to $-0.4$), and the densest filament forms upstream of the minimum-headwind point rather than at the point itself.


\section{Discussion}
\label{sec:dis}
\subsection{Particle drift}

\label{sec:drift}
Section~\ref{sec:res}  showed that T100 forms SI-like filaments inside a
dust-rich, low-headwind region of the bump, while T1 does not, even
though both runs reach favourable local values of $Z/\Pi$. A natural
question is why. Here we argue that the difference is set by the rate
at which particles are delivered through the bump, and that this rate
is itself controlled by the reinforcement timescale $t_{\rm reinf}$.

In a pressure-supported disc, solids feel a headwind from the
sub-Keplerian gas and drift inward through aerodynamic drag. In a
reinforced bump, however, the gas is not evolving freely. The azimuthal gas velocity is relaxed toward a target profile through the term
\begin{equation}
    \left(\frac{\partial u_y}{\partial t}\right)_{\rm reinf}
    =
    -\frac{u_y-\hat{u}_y}{t_{\rm reinf}},
    \label{eq:uy_reinf}
\end{equation}
where $\hat{u}_y$ is the target azimuthal velocity. Because $u_y$ is azimuthal, this term acts as an effective torque on the gas, and a shorter $t_{\rm reinf}$ makes that torque stronger. Our reinforcement scheme also relaxes the gas density towards the target profile on the same timescale $t_{\rm reinf}$ (\autoref{eq:density_reinforcement}), which is not equivalent to torquing the gas. We return to this point in \autoref{sec:caveats}.

The reinforcement acts on the gas, not on the particles.  In the shearing-box particle equations,
\begin{align}
    \frac{\partial V_x}{\partial t}
    &=
    -C\left(V_x-u_x\right)+2\Omega V_y,
    \label{eq:vx_coupling}\\
    \frac{\partial V_y}{\partial t}
    &=
    -C\left(V_y-u_y\right)-\frac{1}{2}\Omega V_x,
    \label{eq:vy_coupling}
\end{align}
where $(V_x,V_y)$ and $(u_x,u_y)$ are the particle and gas velocities, respectively,  and $C$ is the drag coefficient. A change in $u_y$ modifies the azimuthal drag on the particles, and the Coriolis terms transfer that
change into the radial velocity. $t_{\rm reinf}$ therefore does not
apply a radial force on the particles directly. It modifies the
gas--particle drift equilibrium, and the particles respond.

\begin{figure}
    \includegraphics[width=\columnwidth]{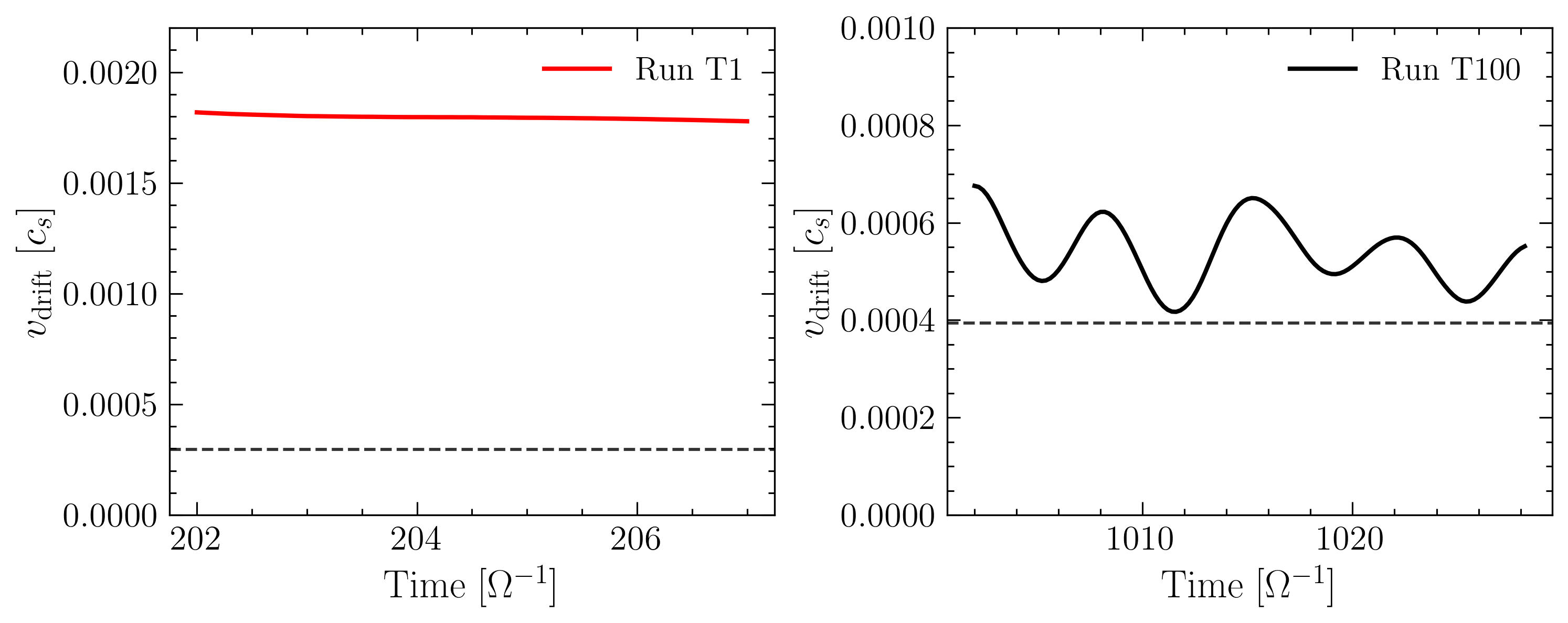}
    
    \caption{Mean radial particle drift velocity (equation~\ref{eq:vdrift_measure}) in the pre-collapse phase. Left: T1 at $t = 202$--$207\,\Omega^{-1}$,  where $v_{\rm drift}\simeq 1.79\times10^{-3}c_{\rm s}$. Right: T100 at  $t = 1000$--$1030\,\Omega^{-1}$, where the mean is $\simeq 5.5\times10^{-4}c_{\rm s}$; oscillations reflect filaments
    drifting through the sampling region. Dashed lines mark the
    Weidenschilling drift speed for the minimum headwind in each run ($\Pi = 0.012$ in T1, $\Pi = 0.016$ in T100). T1
    exceeds the natural rate by a factor of $\sim 6$; T100 sits close to it. Note the different vertical scales.}
    \label{fig:vdrift}
\end{figure}

We measure the outcome directly from the particles. For each output,
we compute the mean inward radial velocity in the midplane bump-feeding region,
\begin{equation}
    v_{\rm drift}
    =
    -\left< v_{p,x}\right>_{\mathcal{R}},
    \qquad
    \mathcal{R}:
    -1.5H \le x \le -0.3H,\quad
    \label{eq:vdrift_measure}
\end{equation}
where the minus sign makes inward drift positive.\footnote{This is a particle-mass-weighted mean, but no masses appear in the expression since all superparticles
carry equal mass} We restrict the measurement to the pre-collapse phase so that it traces the delivery flow rather than gravitationally-bound clump motion. \autoref{fig:vdrift} shows the result. In T1, the pre-collapse mean is
\begin{equation}
    v_{\rm drift,T1} = (1.79 \pm 0.04)\times 10^{-3} c_{\rm s},
\end{equation}
where the uncertainty is the standard deviation across the measurement
window. In T100, the mean is
\begin{equation}
    v_{\rm drift,T100} = (5.5 \pm 1.2)\times 10^{-4} c_{\rm s}.
\end{equation}

The larger scatter in T100 comes from filaments drifting into and out
of the sampling region during the pre-collapse phase. Particles are
delivered through the bump-feeding region roughly three times faster in T1 than in T100. To interpret these values, we compare them with the natural laminar
drift speed. For $\tau_s\ll1$ and $\epsilon\ll1$, where $\epsilon \equiv \rho_p/\rho_g$ is the local particle-to-gas density ratio,
\begin{equation}
    v_{\rm drift,0} \simeq 2\tau_s\,\Pi\, c_s,
    \label{eq:vdrift_natural}
\end{equation}
which for the local values measured across the bump-feeding region gives $v_{\rm drift,0}\simeq3.0\times10^{-4}c_{\rm s}$ in T1 and
$3.9\times10^{-4}c_{\rm s}$ in T100 (\autoref{fig:vdrift}, dashed line in each
panel). T100 exceeds the natural rate by a factor of $\sim1.4$, whereas T1 exceeds it by a factor of $\sim6$. The T100 particles therefore drift close to their natural rate, while the reinforcement torque strongly accelerates the T1 particles.
The relevant quantity for SI filament formation is not the time needed
to cross the full measurement window of equation~\ref{eq:vdrift_measure}, but the residence time inside the narrow high-$Z/\Pi$ region where the SI can grow. We estimate a local crossing time as
\begin{equation}
    t_{\rm cross}
    \sim
    \frac{\Delta x_{\rm SI}}{v_{\rm drift}},
    \label{eq:tcross}
\end{equation}
where $\Delta x_{\rm SI}$ is the width of the favourable region. Taking $\Delta x_{\rm SI}\simeq0.1H$, consistent with the width of the
high-$Z/\Pi$ feature in \autoref{fig:zpi} and \autoref{fig:zpi_T100}, we find
$t_{\rm cross,T1}\simeq 56\,\Omega^{-1}$ and $t_{\rm cross,T100}\simeq 180\,\Omega^{-1}$. These are residence-time
estimates rather than exact trajectories. Their importance is that they can be compared with the SI growth time. Numerical studies find growth times of $t_{\rm grow}\sim100$--$200\,\Omega^{-1}$ for the small-particle regime considered here \citep[e.g.][]{2021ApJ...919..107L}. In T1, the crossing time is shorter than the time for the SI to grow. In T100, it is comparable to $t_{\rm grow}$, giving the instability enough time to reach nonlinear amplitudes before particles are swept out of the high-$Z/\Pi$ region.

This resolves the apparent tension with \citet{2022ApJ...933L..10C}, who proposed the residence-time criterion $t_{\rm cross}>t_{\rm grow}$ but found only GI-driven planetesimal formation in their own simulations. The relevant condition is not whether a bump reaches high $Z/\Pi$, but whether particles remain in the favourable region long enough for SI filaments to grow. T1 violates this condition: particles are delivered quickly into a narrow band that reaches the Roche density and collapses before extended filamentary structure develops. T100 satisfies it: particles drift slowly, remain in the high-$Z/\Pi$ region for longer, and form the SI-like filaments described in \autoref{sec:res}. Both runs are consistent with the same criterion. Changing $t_{\rm reinf}$ changes the angular-momentum exchange with the gas, which changes the particle drift speed, which in turn changes whether the bump acts as a rapid collector or as a long-lived SI-active reservoir.

Two caveats attach to this interpretation. First, our reinforcement scheme is not a physical model of any specific bump-forming mechanism. The Newtonian relaxation applies a uniform torque and a uniform density restoration on a single timescale $t_{\rm reinf}$, whereas a real bump — whether driven by a planet, a zonal flow, or some other process — would exchange angular momentum with the gas through a specific dynamical process with its own spatial and temporal structure. In this sense, $t_{\rm reinf} = 1\,\Omega^{-1}$ is not necessarily a physically realistic timescale. The interpretation that follows from our runs is therefore not that ``strongly reinforced bumps in nature produce GI-driven planetesimals,'' but rather that ``a bump held rigidly by a very strong external torque, in our numerical setup, produces GI-driven planetesimals.''

Second, the density-restoration part of our scheme is more artificial than the torque part. A bump-maintaining process must torque the gas, so the $u_y$ relaxation captures at least the qualitative behaviour of any such mechanism. The $\rho_g$ relaxation, by contrast, has no clear physical analogue: no real process is expected to reset the gas density on a fixed timescale independent of the surrounding dynamics. This part of the scheme becomes particularly aggressive at small $t_{\rm reinf}$, where the midplane gas density is adjusted back to the target profile very rapidly. The T1 result — narrow band, direct GI — may therefore be partly an artefact of this density adjustment rather than a genuine consequence of a strong physical torque. A definitive test of the residence-time picture in strongly reinforced bumps will require simulations with a physically motivated bump-forming mechanism, in which the gas density evolves self-consistently in response to that mechanism.

\subsection{Gas-dynamical effects of $t_{\rm reinf}$}
\label{sec:side_effects_gas}

The reinforcement scheme also affects the gas at the midplane, where the SI operates. We look at this through the total gas kinetic energy, which tracks the total energy in gas motion.

\begin{figure}
    \includegraphics[width=\columnwidth]{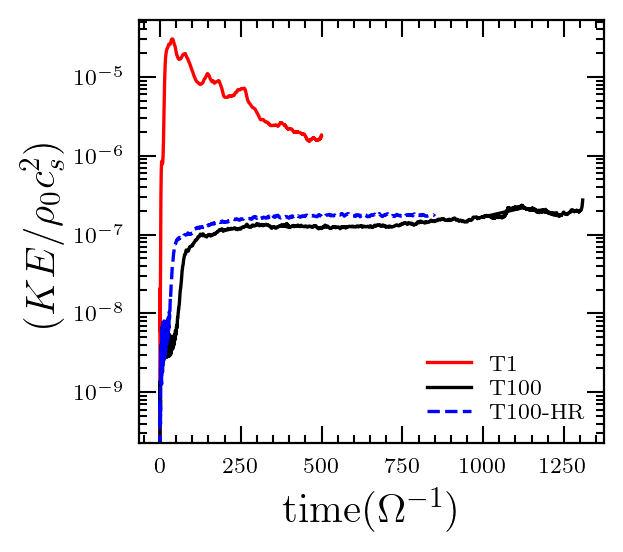}
    \caption{Volume-averaged gas kinetic energy, normalised by $\rho_0 c_s^2$, for T1 (red), T100 (black), and T100-HR (blue dashed). T1 reaches $\sim3\times10^{-5}$ before collapse. T100 settles near $10^{-7}$, two orders of magnitude lower, and T100-HR follows it closely. Note that T1 is run for a shorter duration than the weakly reinforced runs.}
    \label{fig:ke}
\end{figure}

\autoref{fig:ke} shows the kinetic energy. In T1, it rises fast in the
first few tens of orbital times, peaks near $3\times10^{-5}$, and decays by roughly an order of magnitude toward collapse. In T100, it climbs slowly over the first $\sim100\,\Omega^{-1}$ and settles into a long plateau near $10^{-7}$. At peak, that is two orders of magnitude below T1, and T1 remains an order of magnitude above T100 at late times. The midplane gas in T1 is therefore far more energetic than in T100 during the phase when particles are concentrating. T100-HR follows T100 closely throughout, plateauing at a slightly higher level. The quiet midplane in the weakly reinforced runs is therefore not a resolution artefact.

We speculate that this difference arises from the drift itself. Particles in T1 move through the bump-feeding region roughly three times faster than in T100. Only the gas near the midplane is in contact with the particle layer, so faster drift means the particles drag the midplane gas harder while the gas above and below is left alone. This steepens the vertical gradient in the gas velocity, allowing for more turbulent energy to be injected into the system.

Thus, T1 delivers particles into the bump faster \emph{and} into a noisier gas environment. Neither helps the SI. Fast drift shortens the residence time in the favourable region, and midplane gas motion inflates the particle scale height and dilutes the local dust-to-gas ratio \citep[e.g.][]{2018ApJ...868...27Y,2020ApJ...904..132G,2024ApJ...969..130L}. However, the extent to which the increased turbulence suppresses the SI in T1 versus the effect of a shorter residence time on the SI is not known. That said, our analysis of the residence times compared with the growth times of the SI suggests that the residence time is the dominant factor.  A detailed disentangling of these two effects is beyond the scope of this work.

\subsection{Caveats}
\label{sec:caveats}

\textbf{No external turbulence.} The gas is not stirred by MRI, vertical shear instability, convective overstability, or any imposed turbulent forcing. Any velocity fluctuations that develop arise self-consistently from the gas--particle interactions. The choice is deliberate: it isolates the role of the bump reinforcement time in forming planetesimals within the bump. As discussed in \autoref{sec:side_effects_gas}, the midplane in T1 is considerably more agitated than in T100, which we suspect follows from the increased vertical velocity shear driven by the more rapidly drifting particles; we have not verified this directly. Run T100 should therefore be interpreted as a low-external-turbulence limit. In a more turbulent disc, vertical diffusion could thicken the particle layer and reduce the midplane dust-to-gas density ratio, making it harder for the particle layer to reach the densities required for strong SI clumping. Turbulence could also diffuse radially coherent filaments, shorten their lifetime, or fragment them before self-gravity has time to bind the densest regions \citep{2024ApJ...969..130L}.

\textbf{Artificial bump reinforcement.} As discussed in \autoref{sec:drift}, the reinforcement scheme is not a physical model of any specific bump-forming mechanism. It applies a uniform torque and a uniform density restoration on a single timescale, whereas a real bump would exchange angular momentum with the gas through a specific dynamical process. The strongly reinforced run T1 is particularly affected by this caveat, since the density-restoration part of the scheme is most aggressive at small $t_{\rm reinf}$. A definitive test of the residence-time picture in the strongly reinforced regime will require simulations with a physically motivated bump-forming mechanism.

\textbf{Single particle size.} We consider a single particle size corresponding to millimetre grains at 50 au. Real grains in protoplanetary discs span a distribution of sizes set by the balance between coagulation and fragmentation. Multi-species linear theory and non-linear simulations show that the SI behaviour depends sensitively on the number of species and the maximum grain size \citep{2019ApJ...878L..30K, 2021MNRAS.508.5538Y, 2023MNRAS.526.1757R}: relative to the single-size case, size distributions can either enhance or suppress clumping, depending on the maximum grain size and the local dust-to-gas density ratio.

\section{Conclusions}
\label{sec:conclusions}
In this study, we examined whether planetesimal formation can occur with mm-sized grains inside a pressure bump with different reinforcement times. Our main findings are as follows:
\begin{itemize}
    \item The route to collapse strongly depends on the reinforcement time. In the rapidly reinforced T1 run, particles are delivered efficiently into the bump and quickly form a narrow dense band. This band reaches the Roche density at early times and breaks up into gravitationally bound clumps through direct gravitational instability. In the weakly reinforced T100 and T100-HR runs, the growth of the maximum particle density is slower. An SI-like filamentary structure forms inside the pressure bump, where the particles concentrate for several hundred orbits before collapsing and forming gravitationally bound objects. The doubled-resolution T100-HR run reproduces the T100 morphology and collapse sequence, confirming that the SI-driven route in the weakly reinforced regime is not a resolution artefact.
    
    \item The difference between the rapidly and weakly reinforced runs is not only a difference in whether favourable local SI conditions are reached. Both types of run can enter high-\(Z/\Pi\) regions. The key difference is the dynamical history of the particles inside that region. Rapid reinforcement concentrates particles quickly into a narrow band, leading to early direct collapse. Weak reinforcement allows particles to remain in the favourable region for longer, giving SI-like filaments time to develop before self-gravity collapses the densest parts.
    \item The gravitationally bound objects form at different radial locations in T1 and T100. In T1, the compact overdensity forms at the minimum-headwind point, where the bump concentrates particles into a narrow band. In T100, the densest filament forms upstream of the minimum-headwind point, offset by roughly \(0.4H\). The location of gravitational collapse is set by the SI in T100 and by the bump itself in T1. If planetesimals in real discs form through a route similar to T100, they need not form at the exact centre of a dust ring.
    
    \item The reinforcement time also affects the dynamical state of the gas at the midplane. The rapidly reinforced run is more strongly stirred and carries a larger kinetic energy than the weakly reinforced runs. We suspect this follows from the increased vertical velocity shear driven by the more rapidly drifting particles, though we have not verified this directly. In this sense, the reinforcement time affects both the residence time of particles inside the bump and the dynamical state of the gas in which the SI has to grow.
\end{itemize}

\section*{Acknowledgements}

TST and JBS acknowledge support from the NASA Emerging Worlds program under Grant \#80NSSC25K7398. The computations presented in this work were performed on Pleiades through the NASA High-End Computing (HEC) Program at the NASA Advanced Supercomputing (NAS) Division at Ames Research Center under allocation S3005. The simulations were carried out using the \textsc{athena} code.


\section*{Data Availability}
Data will be provided upon reasonable request.



\bibliographystyle{mnras}
\bibliography{example} 





\bsp	
\label{lastpage}
\end{document}